# Protein Photo-folding and Quantum Folding Theory

Liaofu Luo

Faculty of Physical Science and Technology, Inner Mongolia University, Hohhot 010021, China

**Abstract**

The rates of protein folding with photon absorption or emission and the cross section of photon –protein inelastic scattering are calculated from the quantum folding theory by use of standard field-theoretical method. All these protein photo-folding processes are compared with common protein folding without interaction of photons (nonradiative folding). It is demonstrated that there exists a common factor (thermo-averaged overlap integral of vibration wave function, TAOI) for protein folding and protein photo-folding. Based on this finding it is predicted that: 1) the stimulated photo-folding rates show the same temperature dependence as protein folding; 2) the spectral line of electronic transition is broadened to a band which includes abundant vibration spectrum without and with conformational transition and the width of the vibration spectral line is largely reduced; 3) the resonance fluorescence cross section changes with temperature obeying the same law (Luo-Lu's law). The particular form of the folding rate – temperature relation and the abundant spectral structure imply the existence of a set of quantum oscillators in the transition process and these oscillators are mainly of torsion type of low frequency, imply the quantum tunneling between protein conformations does exist in folding and photo-folding processes and the tunneling is rooted deeply in the coherent motion of the conformational-electronic system.

## I  Introduction

Protein is a microscopic system of atoms and molecules. From modern physics it should obey quantum laws in principle. Recently we put forward a protein quantum folding theory [1][2]. Although the bioinformatics studies such as the prediction of protein structure and function from molecular sequence has achieved great successes the dynamical problem on protein folding remains unclear and to be solved. The proposed quantum folding theory emphasizes the idea of torsion cooperative transition. The importance of torsion state can be looked as follows: as a multi-atom system, the conformation of a protein is fully determined by bond lengths, bond angles, and torsion angles (dihedral angles), among which the torsion angles are most easily changed even at room temperature and usually assumed as the main variables of protein conformation. Simultaneously, the torsion potential generally has several minima the transition between which is responsible for the conformational change. All torsion modes between contact residues are taken into account in quantum folding theory. These modes are assumed to participate in the quantum transition cooperatively. About the cooperativeness the Bose condensation of strongly exited longitudinal electric modes of living system was proposed as early as in the seventies of last century [3]. The fold cooperativeness of a protein was also demonstrated in earlier literatures [4]. These authors explained the possible existence of the cooperativeness in protein folding or in living system from the point of non-linear dynamics and thermodynamics. In the meantime, the contact order was introduced as an important parameter for understanding and quantitatively describing folding rate [5]. Recently, the dihedral transition was observed more directly in statistical analysis of protein conformational changes [6]. They indicated the cooperative dihedral transitions occur in most (about 82%) polypeptide chain. Based on above considerations the nonradiative quantum folding theory is formulated. The theory has successfully explained the non-Arrhenius behavior of the temperature dependence of protein folding rates [1][2]. Figure 1 gives an intuitive example, it gives the schematic diagram of the cooperative transition of all torsion angles of protein 1enh (Engrailed Homeo domain).

To explore the fundamental physics behind the folding more deeply and clarify the quantum nature of the folding mechanism more clearly we shall study the protein photo-folding processes, namely, the photon emission or absorption in protein folding and the inelastic scattering of photon on protein (photon-protein resonance Raman scattering). Although the fluorescence technique has been largely developed in recent years and widely employed in studying protein folding and protein-protein interaction dynamics [7][8] the theoretical calculation of fluorescence and other protein photo-folding processes from fundamental interaction seems have not been found in literatures. The situation may be attributed to our "too-classical" understanding of protein folding. But the new-found quantum folding theory affords a sound basis for studying these problems. In the theory the photon emission or absorption in protein folding and the inelastic scattering of photon on protein, as electromagnetic processes, can be accurately described by quantum electrodynamics. The emission or absorption of a photon by the atomic electron is the first step. Then, due to the electronic transition emitting or absorbing photon is coupled to the conformational change of protein structure, the torsion transition in polypeptide chain plays an important role in determining the photon emission /absorption rates or cross sections. We shall make the first-principle-calculation of the rates and cross sections of these processes based on quantum transition theory. The quantitative results provide further checkpoints on the quantum folding theory. It includes: the temperature dependence of the stimulated protein photo-folding rates, the structure and width of the spectral lines of electronic transition in protein-folding, and the temperature dependence of the resonance fluorescence cross section. The experimental tests of these theoretical predictions will support the idea on quantum coherence of the electronic - conformational transition and afford more clear evidences on the quantum nature of protein folding and photo-folding.

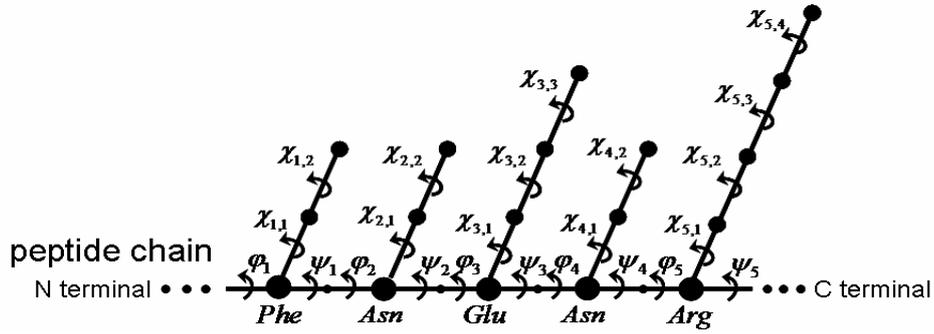

**Figure 1** The schematic diagram of cooperative torsion transition
of protein 1enh (Engrailed Homeo domain)

From the tertiary structure of the protein the 20th amino acid Phe and the 24th amino acid Arg are a pair of contact residues. Five residues Phe, Asn,Glu,Asn and Arg compose a contact fragment. The main chain dihedrals $\varphi_i$ and $\psi_i$ for the *i*-th amino acid and the side chain dihedrals $\chi_{i,j}$ (*j*=1,…4) are labeled in the diagram. $\varphi_i$ means main chain torsion angle for 4 atoms $C_{i-1}, N_i, C^\alpha_i, C_i$; and $\psi_i$ means main chain torsion angle for 4 atoms $N_i, C^\alpha_i, C_i, N_{i+1}$. $\chi_{i,j}$ is defined through side chain 4 atoms in the same manner. The total number of dihedral angles in this example is 23 which are assumed to participate in the quantum transition cooperatively. In protein photo-folding the atomic electron jumps from α to α' emitting or absorbing a photon and the electronic transition is coupled to the torsion transition. **Studying protein photo-folding gives an efficient way to demonstrate the quantum nature of protein folding.**

## 2  Deduction of Protein Photo-folding from Quantum Folding Theory

A protein is regarded as the conformation (torsion coordinate $\{\theta\}$ mainly) – electron system. In adiabatic approximation the wave function of the system can be expressed as

$$M(\theta, x) = \psi(\theta)\varphi(x, \theta) \tag{1}$$

and these two factors satisfy

$$H_2(\theta, x, \nabla)\varphi_\alpha(x, \theta) = \varepsilon^a(\theta)\varphi_\alpha(x, \theta) \tag{2}$$

$$\{H_1(\theta, \frac{\partial}{\partial \theta}) + \varepsilon^\alpha(\theta)\}\psi_{kn\alpha}(\theta) = E_{kn\alpha}\psi_{kn\alpha}(\theta) \tag{3}$$

where $\alpha$ denotes the electron state, and $(k, n)$ refer to the conformational and vibration state, respectively. Since $H_2$ contains electronic kinetic energy term, from gauge invariance of $H_1 + H_2$ we obtain electromagnetic interaction

$$H_{EM} = H_1^{(EM)} + H_2^{(EM)} \tag{4}$$

$$H_1^{(EM)} = -\frac{e}{mc}\sum_{\mathbf{k}\lambda}\frac{\hbar}{\sqrt{2\omega_\mathbf{k}V_0}}(c_{\mathbf{k}\lambda} + c_{\mathbf{k}\lambda}^+)\boldsymbol{\varepsilon}_{\mathbf{k}\lambda} \cdot (-i\nabla) \tag{5}$$

$$H_2^{(EM)} = \frac{e^2}{4mc^2 V_0}\sum_{\mathbf{k}\lambda}\sum_{\mathbf{k}'\lambda'}\frac{\boldsymbol{\varepsilon}_{\mathbf{k}\lambda}\cdot\boldsymbol{\varepsilon}_{\mathbf{k}'\lambda'}}{\sqrt{\omega_\mathbf{k}\omega_{\mathbf{k}'}}}(c_{\mathbf{k}\lambda}c_{\mathbf{k}'\lambda'}^+ + c_{\mathbf{k}\lambda}^+c_{\mathbf{k}'\lambda'} + c_{\mathbf{k}\lambda}c_{\mathbf{k}'\lambda'} + c_{\mathbf{k}\lambda}^+c_{\mathbf{k}'\lambda'}^+) \tag{6}$$

where $m$ is electron mass, $c_{\mathbf{k}\lambda}$ and $c_{\mathbf{k}\lambda}^+$ are annihilation and production operator respectively of photon with wave vector $\mathbf{k}$, frequency $\omega_\mathbf{k}$ and polarization $\boldsymbol{\varepsilon}_{\mathbf{k}\lambda}$ and $V_0$ the normalization volume. From the perturbation $H_1^{(EM)}$ to second order and $H_2^{(EM)}$ to first order we calculate three types of processes, the stimulated single photon emission and absorption in protein folding, the spontaneous photon emission in protein folding and the photon-protein resonance Raman scattering. All calculations are carried out by quantum electrodynamics method. To simplify the notation the calculations are made in units of $\hbar = c = 1$, and only in the final results the Planck constant $\hbar$ and the velocity of light $c$ are written explicitly.

**Stimulated photon emission and absorption in protein folding**

We discuss single photon absorption at first. Set $|i\rangle = |kn\alpha, v_{\mathbf{k}\lambda}\rangle$ where $|kn\alpha\rangle \sim M_{kn\alpha}(\theta, x) = \psi_{kn\alpha}(\theta)\varphi_\alpha(x, \theta)$ and $v_{\mathbf{k}\lambda}$ the photon number of wave vector $\mathbf{k}$ and polarization $\boldsymbol{\varepsilon}_{\mathbf{k}\lambda}$, $|f\rangle = |k'n'\alpha', v_{\mathbf{k}\lambda} - 1\rangle$ where $|k'n'\alpha'\rangle \sim M_{k'n'\alpha'}(\theta, x) = \psi_{k'n'\alpha'}(\theta)\varphi_{\alpha'}(x, \theta)$. For multi-torsion case $\theta = (\theta_1, \theta_2, \ldots \theta_N)$ and $\psi_{kn\alpha}(\theta)$ is the product of the functions of single argument. By using Eq (5) we obtain

$$\langle f|H_1^{(EM)}|i\rangle = -\frac{e}{m}\frac{\sqrt{v_{k\lambda}}}{\sqrt{2\omega_k V_0}}\mathbf{P}_{k'n'\alpha';kn\alpha}\cdot\boldsymbol{\varepsilon}_{k\lambda}$$

$$= -\frac{e}{m}\sqrt{\frac{v_{k\lambda}}{2\omega_k V_0}}\mathbf{P}_{\alpha'\alpha}\cdot\boldsymbol{\varepsilon}_{k\lambda}\int\psi^+_{k'n'\alpha'}(\theta)\psi_{kn\alpha}(\theta)d\theta \quad (7)$$

where

$$\begin{aligned}\mathbf{P}_{k'n'\alpha';kn\alpha} &= \int M^+_{k'n'\alpha'}(\theta,x)(-i\hbar\nabla)M_{kn\alpha}(\theta,x)d\theta dx \\ &= \int\psi^+_{k'n'\alpha'}(\theta)\psi_{kn\alpha}(\theta)d\theta\int\varphi^+_{\alpha'}(x,\theta_0)(-i\hbar\nabla)\varphi_\alpha(x,\theta_0)dx \\ &= \mathbf{P}_{\alpha'\alpha}\int\psi^+_{k'n'\alpha'}(\theta)\psi_{kn\alpha}(\theta)d\theta \end{aligned} \quad (8)$$

$\mathbf{P}_{\alpha'\alpha}$ is the matrix element of electron momentum. In the above deduction of Eq (8) the Condon approximation, namely, the matrix element $\int\varphi^+_{\alpha'}(x,\theta)(-i\hbar\nabla)\varphi_\alpha(x,\theta)dx$ does not dependent on $\theta$ has been used. Assuming the torsion potential (a term occurring in $H_1(\theta,\frac{\partial}{\partial\theta})$ of Eq (3) including $\varepsilon^\alpha(\theta)$ as a part of the potential) has several minima that can be described by a harmonic potential approximately near each minimum the overlap integral $\int\psi^+_{k'n'\alpha'}(\theta)\psi_{kn\alpha}(\theta)d\theta$ of vibration wave function can be calculated [1]. Note that due to $\alpha'\neq\alpha$ the wave function $\psi_{kn\alpha}(\theta)$ and $\psi_{k'n'\alpha'}(\theta)$ are not orthogonal the overlap integral always exists even for $k'=k$. After thermal average over initial vibration states and summation over final vibration states we obtain absorption rate

$$W_a = \frac{\pi e^2}{m^2 V_0}\frac{v_{k\lambda}}{\omega'\omega_k}|\mathbf{P}_{\alpha'\alpha}\cdot\boldsymbol{\varepsilon}_{k\lambda}|^2 I_V \quad (9)$$

$$\begin{aligned}I_V &= \sum_n\left|\int\psi^+_{k'n'\alpha'}(\theta)\psi_{kn\alpha}(\theta)d\theta\right|^2 B(n,T) \\ &= \sum_{\{p_j\}}\prod_j I_{Vj}\end{aligned} \quad (10)$$

$$I_{Vj} = \exp\{-Q_j(2\bar{n}_j+1)\}(\frac{\bar{n}_j+1}{\bar{n}_j})^{p_j/2}J_{p_j}(2Q_j\sqrt{\bar{n}_j(\bar{n}_j+1)}) \quad (11)$$

with

$$\bar{n}_j = (e^{\beta\hbar\omega_j}-1)^{-1} \quad (\beta=\frac{1}{k_B T}) \quad (12)$$

$$Q_j = I_j \omega_j (\delta\theta_j)^2 / 2\hbar \qquad (13)$$

$$p_j = \frac{\delta E_j}{\hbar \omega_j} \qquad (14)$$

$B(n,T)$ is the Boltzmann factor for thermal average, $I_j$ is inertial moment of the $j$-th torsion mode, $\delta\theta_j$ is the angular displacement, and $\delta E_j$ is the energy gap between the initial and final states for the $j$-th mode. $p_j$ represents the net change in quantum number for oscillator mode $j$, which satisfies the constraint

$$\sum_j p_j = p \qquad (15)$$

in the summation of Eq. (10). $I_V$ is called Thermo-Averaged Overlap Integral (TAOI). By use of the asymptotic formula for Bessel function [9]

$$e^{-z} J_p(z) = (2\pi z)^{-1/2} \exp(-p^2/2z) \qquad \text{for} \quad z \gg 1 \qquad (16)$$

$I_{Vj}$ can be simplified. Finally we obtain

$$I_V = \frac{1}{\sqrt{2\pi}} \exp(\frac{\Delta G}{2k_B T})(\sum Z_j)^{-\frac{1}{2}} \exp(-\frac{(\Delta G)^2}{2(\hbar\overline{\omega})^2 \sum Z_j}) \qquad (17)$$

with

$$Z_j = (\delta\theta_j^2) \frac{k_B T}{\hbar^2} I_j \qquad (18)$$

$$\Delta G = \Delta E + \sum_j k_B T \ln\frac{\omega_j}{\omega_j'} \qquad (19)$$

$$(\Delta E = \sum_j \delta E_j)$$

($\overline{\omega}$ is the average of initial torsion frequencies $\omega_j$ over oscillator mode $j$). $\Delta E$ is a potential parameter, the gap of the energy minimum between initial and final torsion state. In Eq (17) the energy gap $\Delta E$ has been replaced by $\Delta G$ to take the frequency difference $\omega_j \neq \omega_j'$ into account. The detailed deduction of $I_V$ can be found in [1]. It is a function of $\omega_j$ (or its average $\overline{\omega}$), $\omega_j'$ (or its average $\overline{\omega}'$), $\delta\theta_j^2$ (or its average $(\delta\theta)^2$) and $\delta E_j$ (or its sum $\Delta E$). Notice that the simplified expression (17) is obtained when $z_j \gg 1$. For photo-folding with conformational change, $k' \neq k$, the condition is fulfilled generally. The single photon absorption cross section is obtained readily from (9)

$$\sigma_a = \frac{e^2}{\hbar c} \frac{\pi}{\overline{\omega}' \omega_\mathbf{k}} \frac{|\mathbf{P}_{\alpha'\alpha} \cdot \boldsymbol{\varepsilon}_{\mathbf{k}\lambda}|^2}{m^2} I_V(\overline{\omega}, \overline{\omega}', \delta\theta, \Delta E) \qquad (20)$$

By comparison with nonradiative folding rate [1]

$$W_{pf} = \frac{2\pi}{\hbar^2 \overline{\omega}'} I_E I_V(\overline{\omega}^{pf}, \overline{\omega}'^{pf}, \delta\theta^{pf}, \Delta E^{pf}) \qquad (21)$$

$$I_E = \frac{\hbar^4}{4} (\sum_j \frac{a_j}{I_j})^2 = \frac{\hbar^4 A}{4 \langle I_j \rangle^2} \qquad a_j = \langle l_j^2 \rangle \qquad A = (\sum_j a_j)^2 \qquad (22)$$

($\langle I_j \rangle$ is the average inertial moment and $l_j$ means the magnetic quantum number of electronic wave function $\varphi_\alpha(x,\theta)$) we obtain the ratio of rates

$$\frac{W_a}{W_{pf}} = 2 \frac{e^2/(\hbar c)}{\hbar^2 m^2 \omega_\mathbf{k}} \frac{\langle I_j \rangle^2}{A} |\mathbf{P}_{\alpha'\alpha} \cdot \boldsymbol{\varepsilon}_{\mathbf{k}\lambda}|^2 \frac{I_V(\overline{\omega}, \overline{\omega}', \delta\theta, \Delta E)}{I_V(\overline{\omega}^{pf}, \overline{\omega}'^{pf}, \delta\theta^{pf}, \Delta E^{pf})} F \qquad (23)$$

$$F = \frac{c v_{\mathbf{k}\lambda}}{V_0} \qquad (24)$$

where $F$ is the photon flux. Setting $\omega_\mathbf{k} \sim 2\pi \times 10^{15}$, $\langle I_j \rangle \sim 10^{-37}$, $(\frac{P}{mc})^2 \sim 10^{-5}$, $A \sim 10 - 10^4$ (all in CGS unit) and $I_V(\overline{\omega}, \overline{\omega}', \delta\theta, \Delta E) = I_V(\overline{\omega}^{pf}, \overline{\omega}'^{pf}, \delta\theta^{pf}, \Delta E^{pf})$ it leads to $\frac{W_a}{W_{pf}} \approx (10^{-23} - 10^{-26}) F$. So, when the photon flux is large enough the single photon absorption rate is comparable with protein folding rate.

The double and multi- photon absorption rates or cross sections can be calculated through the second and the higher order perturbation by the same way. As a general rule they contain the TAOI factor $I_V$.

Set $|i\rangle = |kn\alpha, v_{\mathbf{k}\lambda}\rangle$ and $|f\rangle = |k'n'\alpha', v_{\mathbf{k}\lambda}+1\rangle$ the protein stimulated emission formulas can easily be obtained through calculating matrix element $\langle i|H_1^{(EM)}|f\rangle$. The single-photon stimulated emission cross section is

$$\sigma_e = \frac{\pi e^2}{\hbar c m^2 \overline{\omega'} \omega_\mathbf{k}} \frac{(v_{\mathbf{k}\lambda}+1)}{v_{\mathbf{k}\lambda}} |\mathbf{P}_{\alpha'\alpha} \cdot \mathbf{\varepsilon}_{\mathbf{k}\lambda}|^2 I_V \qquad (v_{\mathbf{k}\lambda} \geq 1) \qquad (25)$$

The double and multi- photon stimulated emission can be calculated through the second and the higher order perturbation. All results contain TAOI factor $I_V$.

**Spontaneous emission in protein folding and the spectral line structure of photo-folding**

Following the same perturbation approach and setting initial $v_{\mathbf{k}\lambda}=0$ in the above deduction of stimulated emission one obtains the single-photon spontaneous emission rate. The rate of quantum transition from a given initial state $|i\rangle = |kn\alpha\rangle$ to the definite final state $|f\rangle = |k'n'\alpha', v_{\mathbf{k}\lambda}\rangle$ is

$$W_{fi} = 2\pi\delta(E_{k'n'\alpha'} + \omega_\mathbf{k} - E_{kn\alpha})|\langle f|H_1^{(EM)}|i\rangle|^2 \qquad (26)$$

Considering that in spontaneous emission case no stimulating electromagnetic field exists and the frequency of emitted photon is not given *a priori*. Adopting continuous representation of electromagnetic field expansion and replacing the sum over photon final states $\frac{\sum_\mathbf{k}}{V_0}$ by $\frac{\int d^3k}{(2\pi)^3}$ we obtain the emission rate or partial width

$$\Gamma_e = \frac{e^2}{\hbar c} \frac{\omega_\mathbf{k}}{2c^2 m^2} |\mathbf{P}_{\alpha'\alpha} \cdot \mathbf{\varepsilon}_{\mathbf{k}\lambda}|^2 I_V$$
$$(\omega_\mathbf{k} = \frac{E_{kn\alpha} - E_{k'n'\alpha'}}{\hbar}) \qquad (27)$$

The numerical estimate gives

$$\Gamma_e \sim 10^8 I_V \sec^{-1} \quad as \quad \omega_k \sim 2\pi \times 10^{15}, (\frac{P}{mc})^2 \sim 10^{-5} \text{ (all in CGS unit)}.$$

As $I_V=1$, Eq (27) is in accordance with Einstein spontaneous emission formulas. We find the spectral line width has been largely reduced due to the overlap integral factor $I_V$. In fact, a spectral line of the electronic transition from state $\alpha$ to $\alpha'$ has been broadened to a band consisting of a lot of single spectral lines. The spectral shape function is determined by the $\delta$ -function $\delta(E_{k'n'\alpha'} + \hbar\omega_\mathbf{k} - E_{kn\alpha})$. For an electronic transition of given frequency $\frac{1}{\hbar}(\varepsilon^{\alpha'}(\theta_0) - \varepsilon^{\alpha}(\theta_0))$ there are a band of spectral lines characterized by the transition $\alpha \to \alpha'$ but with different ($k'n'$) and ($kn$) satisfying $\hbar\omega_\mathbf{k} = E_{kn\alpha} - E_{k'n'\alpha'}$.

**Photon-protein resonance Raman scattering**

For inelastic scattering, set $|i\rangle = |kn\alpha, v_{\mathbf{k}\lambda}, v_{\mathbf{k}'\lambda'}\rangle$, $|f\rangle = |k'n'\alpha', v_{\mathbf{k}\lambda} - 1, v_{\mathbf{k}'\lambda'} + 1\rangle$. The scattering matrix element

$$T_{fi} = \langle f|H_2^{(EM)}|i\rangle + \sum_m \frac{\langle f|H_1^{(EM)}|m\rangle\langle m|H_1^{(EM)}|i\rangle}{E_i - E_m + i0^+} \tag{28}$$

The first term of (28) is

$$\langle f|H_2^{(EM)}|i\rangle = \frac{e^2}{2mV_0} \frac{\boldsymbol{\varepsilon}_{\mathbf{k}\lambda} \cdot \boldsymbol{\varepsilon}_{\mathbf{k}'\lambda'}}{\sqrt{\omega_\mathbf{k}\omega_{\mathbf{k}'}}} \sqrt{v_{\mathbf{k}\lambda}(v_{\mathbf{k}'\lambda'} + 1)} \int M_{k'n'\alpha'}^+(\theta, x) M_{kn\alpha}(\theta, x) d\theta dx$$

$$= \delta_{\alpha'\alpha} \frac{e^2}{2mV_0} \frac{\boldsymbol{\varepsilon}_{\mathbf{k}\lambda} \cdot \boldsymbol{\varepsilon}_{\mathbf{k}'\lambda'}}{\sqrt{\omega_\mathbf{k}\omega_{\mathbf{k}'}}} \sqrt{v_{\mathbf{k}\lambda}(v_{\mathbf{k}'\lambda'} + 1)} \int \psi_{k'n'\alpha'}^+(\theta) \psi_{kn\alpha}(\theta) d\theta \tag{29}$$

Similarly, by using

$$|m\rangle = |k_I n_I \alpha_I, v_{\mathbf{k}\lambda} - 1, v_{\mathbf{k}'\lambda'}\rangle \quad \text{or}$$
$$= |k_I n_I \alpha_I, v_{\mathbf{k}\lambda}, v_{\mathbf{k}'\lambda'} + 1\rangle$$

the second term of (28) is obtained

$$\sum_m \frac{\langle f|H_1^{(EM)}|m\rangle\langle m|H_1^{(EM)}|i\rangle}{E_i - E_m + i0^+} = \frac{e^2}{2m^2 V_0} \frac{\sqrt{v_{\mathbf{k}\lambda}(v_{\mathbf{k}'\lambda'} + 1)}}{\sqrt{\omega_\mathbf{k}\omega_{\mathbf{k}'}}}$$

$$\times \sum_{k_I n_I \alpha_I} \{\frac{(\mathbf{P}_{k'n'\alpha';k_I n_I \alpha_I} \cdot \boldsymbol{\varepsilon}_{\mathbf{k}'\lambda'}) \cdot (\mathbf{P}_{k_I n_I \alpha_I; kn\alpha} \cdot \boldsymbol{\varepsilon}_{\mathbf{k}\lambda})}{E_{kn\alpha} - E_{k_I n_I \alpha_I} + \omega_\mathbf{k} + i0^+} + \frac{(\mathbf{P}_{k'n'\alpha';k_I n_I \alpha_I} \cdot \boldsymbol{\varepsilon}_{\mathbf{k}\lambda}) \cdot (\mathbf{P}_{k_I n_I \alpha_I; kn\alpha} \cdot \boldsymbol{\varepsilon}_{\mathbf{k}'\lambda'})}{E_{kn\alpha} - E_{k_I n_I \alpha_I} - \omega_{\mathbf{k}'} + i0^+}\}$$

(30)

where $\mathbf{P}_{k'n'\alpha';kn\alpha}$ etc can be simplified by Eq(8) under Condon approximation. Considering that $E_{k_I n_I \alpha_I}$ depends on $\alpha_I$ more strongly than $k_I n_I$ and there are a set of resonant intermediate states with same $\alpha_I$ but different $k_I n_I$. The energy of resonant band { $k_I n_I \alpha_{IC}$ } for given $\alpha_I = \alpha_{IC}$ is denoted by $E_{IC}$. Leaving only the resonant term in the summation, near resonance one has

$$\sum_{k_I n_I \alpha_I} \frac{(\mathbf{P}_{k'n'\alpha';k_I n_I \alpha_I} \cdot \boldsymbol{\varepsilon}_{\mathbf{k}'\lambda'}) \cdot (\mathbf{P}_{k_I n_I \alpha_I; kn\alpha} \cdot \boldsymbol{\varepsilon}_{\mathbf{k}\lambda})}{E_{kn\alpha} - E_{k_I n_I \alpha_I} + \omega_\mathbf{k} + i0^+}$$

$$= \int \psi^+_{k'n'\alpha'}(\theta)\psi_{kn\alpha}(\theta)d\theta \frac{(\mathbf{P}_{\alpha'\alpha_I} \cdot \boldsymbol{\varepsilon}_{\mathbf{k}'\lambda'}) \cdot (\mathbf{P}_{\alpha_I\alpha} \cdot \boldsymbol{\varepsilon}_{\mathbf{k}\lambda})}{E_{kn\alpha} - E_{IC} + \omega_{\mathbf{k}} + i0^+} \qquad (31)$$

Similarly,

$$\sum_{k_I n_I \alpha_I} \frac{(\mathbf{P}_{k'n'\alpha';k_I n_I \alpha_I} \cdot \boldsymbol{\varepsilon}_{\mathbf{k}\lambda}) \cdot (\mathbf{P}_{k_I n_I \alpha_I; kn\alpha} \cdot \boldsymbol{\varepsilon}_{\mathbf{k}'\lambda'})}{E_{kn\alpha} - E_{k_I n_I \alpha_I} - \omega_{\mathbf{k}'} + i0^+}$$

$$= \int \psi^+_{k'n'\alpha'}(\theta)\psi_{kn\alpha}(\theta)d\theta \frac{(\mathbf{P}_{\alpha'\alpha_I} \cdot \boldsymbol{\varepsilon}_{\mathbf{k}\lambda}) \cdot (\mathbf{P}_{\alpha_I\alpha} \cdot \boldsymbol{\varepsilon}_{\mathbf{k}'\lambda'})}{E_{kn\alpha} - E_{IC} - \omega_{\mathbf{k}'} + i0^+} \qquad (32)$$

Since $\psi_{kn\alpha}(\theta)$ is the solution of eigenvalue equation (3) we have the completeness of wave functions

$$\sum_{k_I n_I} \psi_{k_I n_I \alpha_{IC}}(\theta)\psi^+_{k_I n_I \alpha_{IC}}(\theta') = \delta(\theta - \theta') \qquad (33)$$

The completeness Eq (33) has been used in the above deduction of Eq (32). Summing up the contribution from $H_1^{(EM)}$ and $H_2^{(EM)}$, inserting equations (29)-(32) into Eq (28) we obtain the cross section of quantum transition from initial state $i$ to a definite final state $f$

$$\sigma_{fi} = \frac{2\pi V_0}{v_{\mathbf{k}\lambda}} \delta(E_{k'n'\alpha'} + \omega_{\mathbf{k}'} - E_{kn\alpha} - \omega_{\mathbf{k}})|T_{fi}|^2 \qquad (34)$$

After thermal average over initial torsion vibration states and summation over final torsion states and photon states in the direction $d\Omega$ (multiplied by a factor $\int d\omega_{\mathbf{k}'} \frac{\omega_{\mathbf{k}'}^2 V_0}{(2\pi)^3}$) we obtain

$$\frac{d\sigma}{d\Omega} = \left(\frac{e^2}{4\pi mc^2}\right)^2 (\nu_{\mathbf{k}'\lambda'} + 1)\frac{\omega_{\mathbf{k}'}}{\omega_{\mathbf{k}}} I_R I_V \qquad (35)$$

where $I_V$ is TAOI given by (17) and

$$I_R = \frac{\left|\frac{1}{m}(\mathbf{P}_{\alpha'\alpha_I} \cdot \boldsymbol{\varepsilon}_{\mathbf{k}'\lambda'}) \cdot (\mathbf{P}_{\alpha_I\alpha} \cdot \boldsymbol{\varepsilon}_{\mathbf{k}\lambda})\right|^2}{(E_{kn\alpha} - E_{IC} + \hbar\omega_{\mathbf{k}})^2 + \frac{\Gamma_{IC}^2}{4}} \qquad (36)$$

or

$$I_R = \frac{\left|\frac{1}{m}(\mathbf{P}_{\alpha'\alpha_I} \cdot \boldsymbol{\varepsilon}_{\mathbf{k}\lambda}) \cdot (\mathbf{P}_{\alpha_I\alpha} \cdot \boldsymbol{\varepsilon}_{\mathbf{k}'\lambda'})\right|^2}{(E_{kn\alpha} - E_{IC} - \hbar\omega_{\mathbf{k}'})^2 + \frac{\Gamma_{IC}^2}{4}} \qquad (37)$$

is Raman tensor and $\Gamma_{IC}$ is the resonance width of $\alpha_{IC}$ band. Eq (35) is in accordance with Kramers-Heisenberg cross section formulas [10] as $I_V = 1$. The factor $\dfrac{e^2}{4\pi mc^2}$ in Eq (35) means electron classical radius. After being excited to a higher quantum state by absorbing photons the orbital electron of a protein can relax to its ground state by emitting a fluorescence photon. The inelastic cross section equation (35) can be used to explain the distribution and polarization of fluorescence photons. The occurrence of TAOI $I_V$ in Eq (35) means the inelastic cross section obeys the same law of temperature dependence as in protein folding.

## 3  Results and Discussions: Test on Protein Quantum Folding Theory

**Common factor of thermo-averaged overlap integral of torsion vibration wave function**
We have studied the protein-photon interaction and deduced the photon absorption/emission cross section and Raman scattering section in protein folding. All these sections (transition rates) have been compared with usual nonradiative folding rates (without interaction of photon). The common features of all photo-protein cross sections are the proportionality of cross section to TAOI $I_V$ of torsion vibration wave function (Eqs (10), (11) and (17)). The factor has also occurred in nonradiative folding rate formulas [1]. It is the generalization of the overlap integral of single mode harmonic oscillators in previous work [11] to the case of multi-modes and non-equal frequencies between initial and final states. Since both initial $\psi_{kn\alpha}$ and final $\psi_{k'n'\alpha'}$ are approximated by the harmonic oscillator wave function, the overlap integral $I_V$ is determined by two sets of harmonic frequencies $\{\omega_j\}$ and $\{\omega'_j\}$ and by the energy gap $\delta E_j$ and angular shift $\delta\theta_j$ between two potentials of the $j$-th torsion modes ($j=1,...,N$). Although the overlapping wave functions in nonradiative folding are $\psi_{kn\alpha}$ and $\psi_{k'n'\alpha}$ with equal quantum number α while in photo-folding are $\psi_{kn\alpha}$ and $\psi_{k'n'\alpha'}$ with $\alpha' \neq \alpha$ both overlap integrals $I_V$'s are same functions of torsion potential parameters $\omega_j$, $\omega'_j$, $\delta\theta_j$ and $\delta E_j$. The analytical form of $I_V$ (Eq.(17)) gives how the transition rate depends on the frequency ratio $\dfrac{\omega_j}{\omega'_j}$ of the $j$-th torsion potential, the potential energy difference $\delta E_j$ and the angular shift $\delta\theta_j$. The overlap integral is classified into two categories: $k = k'$ (without conformational change) and $k \neq k'$ (with conformational change). The protein folding belongs to the second category while the protein photo-folding may occur in both cases, with and without conformational change.

**First test – Temperature dependence of stimulated photon emission and absorption**

The stimulated photon absorption and emission cross sections are given by Eq (20) and (25) respectively. For high incident photon flux the stimulated cross sections are large enough for observation. Since the temperature dependencies of these folding rates and sections are fully determined by factor $I_V$ it leads to a conclusion that they should obey the same temperature relation as nonradiative protein folding. Luo and Lu have deduced a general formula for the temperature dependence of nonradiative transition rate $W$ [2],

$$\frac{d\ln W}{d(\frac{1}{T})} = \frac{d\ln I_V}{d(\frac{1}{T})} = S + \frac{1}{2}T + RT^2 \tag{38}$$

$$S = \frac{\eta \Delta E(T_c)}{2k_B}(1 - \frac{\eta \Delta E(T_c)}{\varepsilon}), \quad R = \frac{k_B}{2\varepsilon}(\lambda + \frac{m}{k_B})^2 \tag{39}$$

where $\varepsilon = \bar{\omega}^2 (\delta\theta)^2 \sum I_j$ is a scale variable of torsion energy, $\delta\theta = \sqrt{<(\delta\theta_j)^2>_{av}}$, $\lambda = \sum_j \ln \frac{\omega_j}{\omega'_j}$ describes the effect deduced from initial-to-final frequency ratio, and $\eta = 1 - \frac{mT_c}{\Delta E(T_c)}$ describes structural susceptibility of torsion potential near melting temperature $T_c$. Now we predict the same temperature dependence (Luo-Lu's law) also holds for photo-folding processes. Since the result is deduced from quantum folding theory it provides a checkpoint for this theory. The experimental test will furnish new evidence on the theory and the view of protein folding as a quantum transition.

Note that in the deduction of temperature dependence, Eqs (38) and (39), the "high temperature approximation", $Z_j = (\delta\theta_j^2)\frac{k_B T}{\hbar^2}I_j >> 1$ has been assumed for Bessel function simplification (see Eq (16)). It requires

$$|\delta\theta_j| > \frac{\hbar}{\sqrt{k_B T I_j}} \tag{40}$$

Fr typical torsion inertial moment $I_j = 10^{-37}$g cm$^2$ it means $\delta\theta_j > 1.6 \times 10^{-2}$ which can be fulfilled for the case of conformational change, $k \neq k'$. However, for the case of photon emission and absorption accompanying small structural relaxation of protein, the conformational change may not occur ($k' = k$) and the angular shift may be small. If Eq (40) is not fulfilled then the original expressions (10) and (11) for TAOI should be used instead of Eq (17). In this case, the temperature dependence is more complicate than given by Eqs (38) and (39).

**Second test –Broadening and structure of electronic transition spectrum**

The motion of orbital electron obeys wave equation Eq (2). In a given macromolecular configuration $\theta=\theta_0$ the energy is $\varepsilon^\alpha(\theta_0)$ and the emitted photon frequency is $\omega_\mathbf{k} = \frac{1}{\hbar}(\varepsilon^\alpha(\theta_0) - \varepsilon^{\alpha'}(\theta_0))$ as the electron jumping from $\alpha$ to $\alpha$'. However, due to the coupling between protein structure and electron motion the electronic jump inevitably causes protein structural relaxation or conformational change. That is, the quantum state of the conformation-electron system changes from $M_{kn\alpha}(\theta,x)$ to $M_{k'n'\alpha'}(\theta,x)$ due to electronic transition. The protein structure variation, in turn, makes the frequency of emitted photon shifting from $\omega_\mathbf{k}$ to $\omega_\mathbf{k} + \Delta\omega_\mathbf{k}$ with

$$\Delta\omega_\mathbf{k} = \frac{1}{\hbar}(E_{kn\alpha} - E_{k'n'\alpha'}) - \frac{1}{\hbar}(\varepsilon^\alpha(\theta_0) - \varepsilon^{\alpha'}(\theta_0)) \qquad (41)$$

Thus the electronic transition spectrum is broadened and a spectral band is formed corresponding to electronic transition $\alpha \to \alpha$'. The width of the spectral band is determined by the torsion vibration frequency. For example, for the spectral line $\omega_\mathbf{k} \sim 2\pi \times 10^{15}$ sec$^{-1}$, the band width is in the order of $10^{13}$ sec$^{-1}$, one hundredth or thousandth of the frequency, and consists of transitions between $\alpha$ and $\alpha$' among several tens of vibration energy levels.

Now we discuss the width of each spectral line in the band. The rate of spontaneous emission of photon in protein folding is given by Eq (27). It contains the overlap integral factor $I_V$. The reason can be explained by the following argument. When an electron jumps from one atomic orbital to another in the same molecular harmonic potential the transition obeys the strong selection rule. All transitions with changing vibration quantum number will be forbidden due to the orthogonality of wave functions. However, for an electronic transition with initial and final torsion vibration in different harmonic potentials the vibration wave functions $\psi_{kn\alpha}(\theta)$ and $\psi_{k'n'\alpha'}(\theta)$ cannot be orthogonal to each other and the overlap integral exists. This is so-called 'forbidden' transition. The overlap integral TAOI is an important determinant factor of the 'forbidden' transition rate. In the preceding calculation of single-photon emission we estimate $\Gamma_e \sim 10^8 I_V$ sec$^{-1}$. $I_V$ changes in a wide range. From Eq (17) (18) one has

$$\frac{d \ln I_V}{d(\delta\theta)^2} = -\frac{1}{2}\frac{1}{(\delta\theta)^2} + \frac{(\Delta G)^2}{2N\langle I_j\rangle \overline{\omega}^2 k_B T}\frac{1}{(\delta\theta)^4} \qquad (42)$$

where $N$ = the number of collective torsion modes of the polypeptide chain. It gives $I_V$ taking maximum at

$$(\delta\theta)^2_{max} = \frac{(\Delta G)^2}{N\langle I_j \rangle \bar{\omega}^2 k_B T} \tag{43}$$

Setting $\Delta G$ = -2Kcal/mol (-4Kcal/mol), $N\langle I_j \rangle$ = $10^{-35}$g cm$^2$, $\bar{\omega}=10^{12}$ sec$^{-1}$ one estimates $(\delta\theta)_{max}$ =0.21 (0.42) and the corresponding $(I_V)_{max}=3.2\times 10^{-4}(2.9\times 10^{-5})$. $I_V$ increases with $\delta\theta$ due to the second term of Eq (42) and attains the maximum, then decreases with $\delta\theta$ due to the first term of it. In the left of maximum $I_V$ rapidly decreases as $\delta\theta \to 0$ following $\exp\{-\frac{const}{(\delta\theta)^2}\}$. In the right of maximum $I_V$ changes with $(\delta\theta)^{-1}$ approximately. Assuming $I_V \sim 10^{-5}$ we obtain $\Gamma_e \sim 10^3$ sec$^{-1}$. A typical lifetime for an atomic energy state is about $10^{-8}$ seconds, corresponding to a natural linewidth of about 6.6 × $10^{-8}$ eV. So the width $\Gamma_e$ of the spectral line in protein photo-folding is five orders smaller than the natural linewidth. In fact, due to its exponential dependence on $(\delta\theta)^2$, $I_V$ may take a value much lower than $10^{-5}$ for small $\delta\theta$. It leads to the extra-narrowness of the width $\Gamma_e$. This is a well-marked characteristic of the photo-folding spectral lines.

**Third test – Temperature dependence of resonance fluorescence cross section**

The cross section of inelastic photon-protein resonance Raman scattering is given by Eqs (35)-(37). Due to the TAOI factor $I_V$ the inelastic cross section behaves the same temperature dependence as in protein foding rates. Non-Arrhenius behavior, Eq (38), should be observed in experiments.

The general relation Eq (38) of temperature dependence is a characteristic of torsion quantum oscillators. To clarify this point we consider the vibration of bond length and bond angle (stretching and bending) for comparison. Of course, both the bond length and the bond angle of macromolecule are important dynamical variables and their vibration can be put into the formalism of quantum theory. However, their temperature dependence should exhibit another characteristics different from Eq (38). The reason is twofold. First, due to the frequency difference between two kinds of modes − torsion and stretch /bend, the Boson condensation occurs only in torsion but not generally in stretch or bend modes. Consider the vibration partition function of a molecule in harmonic conformational potential.

$$Z = (e^{(1/2)\beta\hbar\omega} - e^{-(1/2)\beta\hbar\omega})^{-1} \tag{44}$$

The entropy $S$ can readily be deduced,

$$S = k_B\{\beta\hbar\omega(e^{\beta\hbar\omega} - 1)^{-1} - \ln(e^{\beta\hbar\omega} - 1) + \beta\hbar\omega\} \tag{45}$$

By means of (45) one finds the strong dependence of $S$ on frequency $\omega$. For example, at $T$ =

300 K as the frequency $\nu = \frac{\omega}{2\pi}$ is lowered down to $10^{13}$ Hz, the entropy increases rapidly with the decrease of $\nu$,

$$S = (k_B \ln 2) \times 10^{-6}, \quad (k_B \ln 2) \times 0.63, \quad (k_B \ln 2) \times 2.83 \quad \text{at} \quad \nu = 10^{14}, 10^{13}, 10^{12} \text{ Hz},$$

respectively. So, the Boson condensation occurs and the free energy of the oscillator system drops for torsion vibrations of low frequency. There is a large gap of free energy between two kinds of oscillators. Second, for bond stretching or bending, due to lack of multi-minima in the potential only the protein structural relaxation can occur in electronic transition and no large conformational change can be defined. So the temperature dependence of stretching and bending oscillation should exhibit characteristics different from Eq (38).

**Remarks on protein folding without interaction with photon**

The comparative study of protein photo-folding and nonradiative folding can give important information on folding mechanism. The ratio of protein photo-folding rate to corresponding radiation-less folding takes a relatively simple form if the overlap integrals $I_V$ of two processes are approximately equal and can be canceled in the ratio (Eq 23). Apart from the overlap integral $I_V$ of conformational wave function the nonradiative folding rate is dependent of the matrix element $I_E$ of electronic wave function. Only in case of the quantum number $\alpha$ equal $\alpha'$, $I_E$ can be simplified and formally expressed by $A$ (Eq (22)). However, the matrix element of electronic wave function in photo-folding is more easily deduced and has been expressed in a simple form $\mathbf{P}_{\alpha'\alpha} \cdot \boldsymbol{\varepsilon}_{\mathbf{k}\lambda}$ (Eqs (9), (20), (25)). This is an advantage of protein photo-folding in theoretical studies. Thus, the comparative study of radiation-less folding and photo-folding can give more information on protein folding, including the information on the electronic wave function factor $I_E$ (or $A$) of the nonradiative folding.

Although the condition of $\alpha = \alpha'$ holds for most nonradiative folding, how the result changes if the condition is not assumed? The general form of nonradiative transition matrix element is given by [1]

$$\langle k'n'\alpha' | H' | kn\alpha \rangle = \int \psi^+_{k'n'\alpha'}(\theta) \{-\frac{\hbar^2}{2I} \int \varphi^+_{\alpha'} (\frac{\partial^2 \varphi_\alpha}{\partial \theta^2} + 2\frac{\partial \varphi_\alpha}{\partial \theta} \frac{\partial}{\partial \theta}) d^3x\} \psi_{kn\alpha}(\theta) d\theta$$

(46)

Apart from the first term which has been considered previously in [1][2]. The second term of Eq (44) is a new term which can be rewritten in

$$\langle k'n'\alpha' | H'_{new} | kn\alpha \rangle = \int d\theta \{ -\frac{\hbar^2}{I} \int \varphi_{\alpha'}^+ \frac{\partial \varphi_\alpha}{\partial \theta} d^3x \} \psi_{k'n'\alpha'}^+(\theta) \frac{\partial}{\partial \theta} \psi_{kn\alpha}(\theta)$$
$$= -\frac{\hbar^2}{I} \int \varphi_{\alpha'}^+ \frac{\partial \varphi_\alpha}{\partial \theta} d^3x \Big|_{\theta=\theta_0} \int \psi_{k'n'\alpha'}^+(\theta) \frac{\partial}{\partial \theta} \psi_{kn\alpha}(\theta) d\theta \quad (47)$$

Here, the first integral of Eq (47) is calculated at the point $\theta_0$ of maximum overlap of wave function $\psi_{kn\alpha}(\theta)$ and $\partial_\theta \psi_{k'n'\alpha'}(\theta)$. The thermal average of the overlap integral occurring in the transition rate is

$$I_{V,new} = \sum_n \left| \int \psi_{k'n'\alpha'}^+(\theta) \frac{\partial}{\partial \theta} \psi_{kn\alpha}(\theta) d\theta \right|^2 B(n,T) \quad (48)$$

Therefore, the temperature dependence of the folding rate is determined by both $\frac{d \ln I_V}{d(\frac{1}{T})}$ and $\frac{d \ln I_{V,new}}{d(\frac{1}{T})}$. However, $I_{V,new}$ was discussed only for single mode case in the past literatures [12][13][14]. More detailed studies are needed for the complete solution of the problem.

**Conclusions and prospective**
The particular form of the same temperature dependence (Eq 38) for protein nonradiative folding and photo-folding and the abundant structure of the photo-folding spectral band consisting of many narrow lines are two main results deduced from protein quantum folding theory. These results are closely related to the fundamental concepts of quantum mechanics. First, they imply the existence of a set of quantum oscillators in the transition process and these oscillators are mainly of torsion vibration type of low frequency. Second, they imply in protein folding the quantum tunneling does exist which means the non-locality of state and the quantum coherence of conformational-electronic motion. The coherence is rooted deeply in the cooperative motion of many structural constituents (atomic electrons, molecular torsion, etc) under given temperature. More experimental tests on the above predictions are waited for. It will give more evidences on the quantum nature of protein folding and photo-folding.

Due to the separability of conformational motion and electronic motion the basic problem of protein quantum folding can be formulated in a simple form as follows: to calculate the quantum tunneling of a set of quantum oscillator jumping from one conformation to another under given temperature. We have demonstrated that the problem can be solved in TAOI (thermo-averaged overlap integral of vibration wave function) formalism. With the aid of it not only the transition between two conformations can be calculated, but also the tunneling between molecules in a larger system can be studied as well. It is expected that the quantum motion exists not only in a protein as discussed in this article but also in a gene, in a larger system of cell and even in human brain. Moreover, the oscillators (molecules) that are excited through light absorption or via a different process can transfer energy to a second 'sensitized' oscillator (molecule), which is converted to its excited state and can then propagate the excitation further. TAOI as a measure of the correlation of two quantum oscillators can be introduced in this problem and the subject about the long-distance energy and information transfer between microscopic constituents of living system can be studied in this approach.

**Acknowledgement** The author is indebt to Drs Jun Lu and Ying Zhang for numerous discussions on the experimental data analysis of protein folding rate.